\documentclass[showpacs,aps,twocolumn,prl,amssymb]{revtex4}
\usepackage{dcolumn}
\begin{document}

\textbf{Comment on ``Critical State Theory for Nonparallel Flux
Line Lattices in Type-II Superconductors''}

Recently, Bad\'ia and L\'opez (BL) proposed a new model relevant
to the electrodynamics of hard superconductors in multicomponent
situation~\cite{bad}. BL have claimed that using their least
action principle with the given constrains on the current density
one can solve any critical state problem. They argue that the BL
model is equivalent to the set of Maxwell equations with the
vertical current-voltage characteristics (CVC). However, for the
isotropic case (the only case analyzed in~\cite{bad}) the BL
formalism does not lead to new results in comparison to the Bean
critical state model generalized for the multicomponent case. As
regards the anisotropic situation, we show here that the BL model
\textit{is not equivalent to the set of Maxwell equations with the
``vertical CVC''}. In other words, we call in question the
validity of the BL model in a general case.

BL proposed a new formalism for solving the critical state
problems and applied it for the analysis of the isotropic
situation. For this particular case, the BL formalism is surely
adequate and is completely equivalent to the set of Maxwell
equations with the CVC in the form, ${\vec J}=J_c {\vec E}/E$,
where $J_c$ is the critical current density, ${\vec E}$ is the
electric field. This material equation corresponds to the
generalized Bean critical state model (GBCSM). The GBCSM is
commonly used in the electrodynamics of hard superconductors.
Specifically, the analysis of this model allowed us to discover a
new interesting phenomenon of the collapse of the transport
current~\cite{col} and of the magnetization~\cite{buf}. Contrary
to~\cite{bad}, the GBCSM is useful for clarification the nature of
the collapse. Indeed, let us consider a dissipation-free dc
critical current ${\vec J}_{dc}$ shielding external dc magnetic
field ${\vec H}$ and providing dc magnetization $M$ of the
superconductor. After switching on ac magnetic field ${\vec
h}\perp {\vec H}$, ac current ${\vec J}_{ac}$ is induced (${\vec
J}_{ac} \perp {\vec J}_{dc}$). According to GBCSM, the total
current should flow along ${\vec E}$. This means that the
dissipation-free current $J_{dc}$ is suppressed and the collapse
of $M$ appears to be. It is natural that owing to equivalence of
the BL and GBCSM models, simulations concerning the collapse
in~\cite{bad} should coincide with the results obtained previously
(see e.\ g.\ \cite{colnew}).

Although Letter~\cite{bad} deals with an analysis of the isotropic
case, BL declare that ``\textit{several extensions may be
implemented if dictated by the physics of the problem. These
include the effect of equilibrium magnetization by means of an
appropriate $B(H)$ relation, the selection of the model $J_c(H)$
and the use of anisotropic control spaces}''. It seems to us that
the validity of the BL formalism in the anisotropic case is
questionable. Although the BL model operates correctly with the
region of the dissipation-free current flow, we argue that the set
of Maxwell equations with CVC is not equivalent to the BL model
for the dissipative region. Indeed, CVC of samples in the critical
state contains not only information about a region where the
current flow is dissipation-free, \textit{but provides the
one-to-one correspondence between the directions of the critical
current and the electric field within the dissipative region}. It
should be noted that not only a single CVC but a class of CVC can
correspond to the same restriction on the current density. For
example, the class of CVC with an arbitrary parameter $\alpha$,
\begin{equation}\label{1}
  J_x=J_c \cos(\theta +\alpha), \qquad J_y=J_c\sin(\theta +\alpha),
\end{equation}
corresponds to the isotropic constraint $|{\vec J}|\le J_c$ but
describes the case where the direction of ${\vec J}$ is inclined
with respect to ${\vec E}$. Here, $x$ and $y$ are arbitrary
coordinate axes in the sample plane, $\theta$ is the angle between
the vector ${\vec E}$ and $x$ axis. Obviously, the set of Maxwell
equations together with these CVC would give different profiles
for the field penetration, whereas the BL formalism with the
isotropic constraint gives a unique solution of the problem. This
means, that the variational procedure chooses only one of CVC from
a class, ignoring the real CVC. This choice is defined by the
requirement of the minimum of the BL functional. Only for the
isotropic case such a requirement predetermines $\alpha =0$ in
Eq.~(\ref{1}) (i.\ e.\ ${\vec J}\|{\vec E}$) owing to the symmetry
of the problem. In other cases, one cannot be sure that the
suggested least action principle chooses the CVC correctly. Of
course, if this principle is proved, it would be a very important
tool to determine the CVC via the known restriction region.
Unfortunately, there is no reason to expect this because the
minimum principle for the entropy production rate is not proved
for nonlinear unsteady systems.

Thus, the BL formalism being valid for the isotropic case can
hardly be correct for more complex situations, in particular, for
anisotropic cases.

This work is supported by INTAS and RFBR, projects IR-97-1394 and
00-02-17145.

Leonid M. Fisher$^1$ and Valery A. Yampol'skii$^2$

$^1$All-Russian Electrical Engineering Institute, 12
Krasnokazarmennaya Street, 111250 Moscow, Russia,

$^2$ Institute for Radiophysics and Electronics Ukrainian Academy
of Science, 12 Proskura Street, 61085 Kharkov, Ukraine


\end{document}